# Is prokaryotic complexity limited by accelerated growth in regulatory overhead?

Larry J. Croft[1]*, Martin J. Lercher[2], Michael J. Gagen[1,3], and John S. Mattick[1]

[1] ARC Special Research Centre for Functional and Applied Genomics, Institute for Molecular Bioscience, University of Queensland, Brisbane, Qld 4072, Australia; [2] Department of Biology & Biochemistry, University of Bath, Bath BA27AY, UK; [3] Physics Department, University of Queensland, Brisbane, Qld 4072, Australia; * Present address: Novartis Institute for Biomedical Research, Basel 4056, Switzerland; Contact author email: j.mattick@imb.uq.edu.au

**Increased biological complexity is generally associated with the addition of new genetic information, which must be integrated into the existing regulatory network that operates within the cell. General arguments on network control, as well as several recent genomic observations, indicate that regulatory gene number grows disproportionally fast with increasing genome size. We present two models for the growth of regulatory networks. Both predict that the number of transcriptional regulators will scale quadratically with total gene number. This appears to be in good quantitative agreement with genomic data from 89 fully sequenced prokaryotes. Moreover, the empirical curve predicts that any new non-regulatory gene will be accompanied by more than one additional regulator beyond a genome size of about 20,000 genes, within a factor of two of the observed ceiling. Our analysis places transcriptional regulatory networks in the class of accelerating networks. We suggest that prokaryotic complexity may have been limited throughout evolution by regulatory overhead, and conversely that complex eukaryotes must have bypassed this constraint by novel strategies.**

## Background

An increase in organism complexity or functionality can be achieved by adding new functional genes, and/or by adding new regulatory regimes. Each case requires an expansion of the regulatory network to integrate new capabilities with existing ones. The ability to access more genes (or operons, i.e., co-regulated functional modules of genes) therefore not only involves a linear increase in regulator number, but also enforces an additional expansion of higher order regulation, if the system as a whole is to be coordinated and not descend into chaotic space. A fraction of the regulators (those that are not constitutively expressed) will themselves require regulation, if the system as a whole is to be coordinated and not descend into chaotic space. A fraction of the regulators (those that are not constitutively expressed) will themselves require regulation, and the impact of new gene products on the biology of the cell will need to be integrated by additional input into the existing regulatory framework. For example, if a new operon dealing with the metabolism of a particular sugar is introduced into the cell, not only is a new regulator that recognizes this sugar required (or at least advantageous), but the effect of the activity of this operon has to be coordinated with the metabolism of other substrates that feed into the cell's energy flux, as exemplified by the *lac* operon. A growing body of literature supports the notion that increases in complexity arise indeed from progressively more elaborate regulation of gene expression [1]. These considerations suggest that the numbers of regulators (or combinations thereof) must generally scale faster than linearly with the number of genes.

In agreement with this general prediction, it has been shown that regulatory gene number in prokaryotic genomes grows disproportionally fast [2-5]. In particular, a recent study analyzed the scaling of gene counts $n_c$ for each of 44 functional protein categories in relation to the total number of genes $n$, across 64 bacterial genomes [5]. Surprisingly, almost all categories showed a power law dependence on total gene count, $n_c \sim n^\alpha$. Transcriptional regulators were the fastest growing class, with an exponent $\alpha$ of approximately 2 (1.87±0.13 for "transcription regulation" and 2.07±0.21 for "two component systems"). As linear increases in regulator numbers can theoretically provide a combinatorially explosive number of regulatory regimes, this observation suggests that the number of required regulatory states is increasing faster than the number of meaningful combinations of regulatory factors [5], although the upper limits on the size of genetic networks that may be imposed by this regulatory expansion were not considered.

Studies of such 'accelerating growth' networks have recently been prompted by observations that the Internet grows by adding links more quickly than sites [6]. However, the relative change over time is small and the Internet appears to remain scale free and well characterized by stationary statistics [7]. Similarly, the average number of links per substrate in metabolic networks of organisms appears to increase linearly with substrate number [8], while the average number of links of scientific collaboration networks increases linearly over time [9]. These observations have motivated models where accelerating growth in link number generates nonstationary statistics from random to scale-free to regular connectivity at particular network sizes, as the growing number of links gradually saturates the network and links all nodes together [10] (for an overview, see [11, 12]). If biological regulatory networks indeed feature accelerating growth, they will be characterized by sparse connectivity at low gene numbers. If these networks, optimized by evolution in the sparse regime, are unable to make the transition to the densely connected regime, the evolutionary record would show a strict size limit at some maximum network size. This is exactly what is observed: prokaryotic gene numbers appear restricted to below approximately 10,000 genes or a genome size of approximately 10 megabases [13].

Below, we present mathematical arguments that substantiate our intuitive expectation of accelerated growth of regulatory networks; and we confirm that our predictions are in good agreement with experimental results. We further calculate a rough estimate of the size limit that this accelerated growth of regulatory networks may impose on prokaryotic genomes.

We wish to point out that both models outlined below rely on a number of plausible, yet unproven assumptions. These models will undoubtedly need adjustments once more detailed information on the evolutionary mechanism of genome expansion is available. While they will not encapsulate all biological details correctly, the models may serve as illustrations of the kind of mechanisms that might cause the observed accelerating growth of regulatory networks.

# Results and Discussion

## Model I: Regulatory integration of new genes

We can derive a specific prediction of the relationship between regulator numbers $R$ and the numbers of genes $N$ from simple assumptions about the evolution of regulatory networks. Consider a new gene that is added to the genome, e.g., by gene duplication or horizontal gene transfer. Initially, the new gene will be free to drift in its pattern of interactions with other genes. The available space of regulatory interactions that it can explore is the full set of genes already present in the genome. This evolutionary search is undirected. Thus, *a priori*, each regulatory interaction of the new gene with any previous gene has the same probability (termed $p$) to be selectively favorable. For each gene added to a genome containing $N$ genes, we thus expect $p\,N$ interactions to become fixed. Some genes may be integrated into the regulatory network only via existing regulatory factors. However, we expect that some of the new genes have to be regulated specifically; thus, a fraction $v$ of the new interactions will correspond to new regulatory factors. In sum, adding one new gene results in the fixation of $\Delta R = v\,p\,N$ new regulators, with $v\,p = c$ constant; or equivalently, adding $\Delta N$ new genes results in $\Delta R = c\,N\,\Delta N$ new regulators. Some (but perhaps only a minority) of the $\Delta R$ new regulators will themselves require regulation [14], and reciprocally the activities of the newly regulated functional units will have to be integrated back into the regulatory network of the cell, both leading to additional higher order terms dependent on the degree of required connectivity of the system as a whole. In a first approximation, we will ignore these; their inclusion will further accelerate regulatory network growth.

Starting from a hypothetical empty genome and adding one gene at a time, we can estimate the total number of regulators as a sum over all $\Delta R$ terms:

$$R = \sum_{n=0}^{N} cn = \frac{cN(N+1)}{2} \approx \frac{c}{2}N^2 \qquad (1)$$

Thus, the number of regulators $R$ scales approximately quadratically with increasing gene number. As prokaryotic operon size decreases only slowly with increasing genome size [15], $N$ in Eq.1 can also be interpreted as operon number, which simply changes the scaling factor $c$. In eukaryotes, $N$ includes the numbers of different splice variants.

## Model II: Homology based interactions of new regulators

An alternative theoretical approach focuses not on the regulation required by newly added genes, but on the transcriptional regulators themselves. Any given transcription factor, which is newly added to a genome, will be retained under one condition: that it establishes fitness enhancing interaction with potential binding sites present in the genome.

We assume that the nucleotide sequences of potential binding sites are approximately random. The probability of finding or developing a match to the given transcription factor specificity among potential transcription factor binding sequences is then proportional to the total amount of such sequences. This scaling is analogous to that known from sequence similarity searches, such as BLAST or FASTA: the probability of finding a random sequence match scales linearly with the length of the target sequence.

The total amount of potential transcription factor binding sequences will scale linearly with the number of genes. Thus, the average number of matches between our new transcription factor and potential binding sequences scales linearly with total gene number. If the evolutionary search is undirected, i.e., each interaction is *a priori* equally likely to provide a fitness benefit, then the probability of retaining a newly added transcription factor also scales linearly with gene number: $\Delta R = c\,'N$, again leading to Equation (1).

## Genomic analysis

We then proceeded to examine the actual relationship between the numbers of transcriptional regulators (defined here as those utilizing sequence specific binding to DNA or RNA) and genome size in prokaryotes. We analysed 89 completely sequenced bacterial and archaeal genomes, ranging from *Mycoplasma genitalium* (containing just 480 genes) to *Bradyrhizobium japonicum* (8317 genes), thereby spanning almost the entire range of observed genome sizes. For each genome, we estimated the number of regulatory proteins by searching all genes for matches to Pfam profiles of protein domains [16] with known regulatory or signalling functions and/or known to be involved in DNA or RNA binding. Although all of these genomes contain many genes with unknown function, some of which may have potentially unidentified regulatory domains, this should not unduly bias the analysis unless the proportion of unidentified regulators varies with genome size, which appears unlikely.

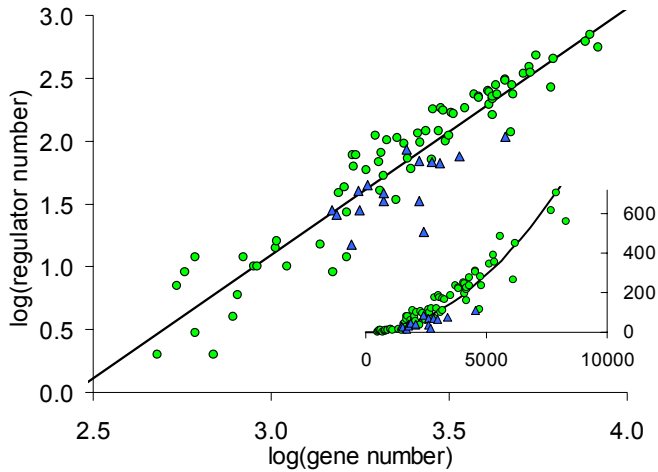

**Figure 1:** Double-logarithmic plot of transcriptional regulator number against total gene number for bacteria (green circles) and archaea (blue triangles). The overall distribution is well described by a straight line with slope 1.96 ($r^2$= 0.88, 95% confidence interval: 1.81 – 2.11), corresponding to a quadratic relationship between regulator number and genome size. The inset shows the same data before log-transformation.

In agreement with the predictions from the above models, and in agreement with previous studies [2, 5], we find the increase of regulatory genes with total gene number to be most consistent with a quadratic function: the double logarithmic plot in Fig. 1 is well described by a straight line with slope 1.96 (95%-confidence interval 1.81 - 2.11, $r^2$=0.88), giving the empirical relationship $R = 0.0000163\ N^{1.96}$. The fitted curve is not significantly different from $R = 0.0000120\ N^2$. Importantly, the relationship remains consistent with a quadratic even when more or less stringent definitions of regulatory protein domains are employed, or when all proteins annotated as regulators are included (data not shown; see also [5]). While the size range of fully sequenced prokaryotic genomes spans hardly more than one order of magnitude, the high $r^2$ value together with the tight confidence interval for the exponent are good evidence that the true scaling behaviour is a higher order function (i.e. higher than linear), and most likely to be close to a quadratic.

From Fig. 1, it is evident that regulatory networks are sparsely connected. In *E. coli*, each transcriptional regulator targets on average 5 operons [17], corresponding to approximately 8.5 genes [15]. If this relation can be extrapolated to other genomes, Fig. 1 suggests that each gene is connected to on average $C = 0.06\%$ of all other genes (see Methods).

Previous studies of the transcriptional network in *Escherichia coli* have found a modular structure [17, 18]: densely co-regulated sets of genes form partially overlapping functional modules, which are controlled by global regulators. If new genes explore regulatory interactions predominantly within modules that are sparsely connected into the rest of the network, then this would give a largely linear relationship between $R$ and $N$, with a small quadratic term for module interconnectivity. The fact that the relationship between $R$ and $N$ appears to be close to a pure quadratic (Fig. 1) then suggests that new genes explore regulatory interactions with a significant proportion of the genome, not just within modules. This is consistent with the finding that modules do not represent closed systems: many genes are regulated not only by within-module transcription factors, but also by factors that control genes across several modules [17].

## Regulatory networks

The observed increase of average link number with network size means that transcription networks feature accelerating growth in connectivity, and hence nonstationary (or size dependent) statistics. This constitutes a significant difference from the more usual classes of exponential or scale-free networks with stationary (size independent) statistics. The non-linear relationship predicted in Eq.1 and confirmed in Fig. 1 differs from the growth behavior described for previously studied metabolic and man-made networks, which largely feature non-accelerating growth and stationary statistics [6, 8, 9, 19, 20]. This may be a consequence of fundamental differences: the Internet, the World Wide Web, and scientific collaboration networks, among other generally studied networks, have not been subject to selection for specific dynamical functions, as opposed to gene regulatory networks [21]. While metabolic networks are of course related to the regulatory networks governing them (and are indeed optimized for a closely related function), they are dominated by the most highly connected substrates (such as water, ATP, and ADP) [8]; links involving such ubiquitous reactants contain little information on network control.

The scaling law in Eq. 1 is based on two simple suppositions: that each new gene explores a space of possible interactions which is proportional in size to the total number of genes; and that *a priori* each new interaction is equally likely to lead to the fixation of a new regulator. We can develop this into a simple explicit network model by presuming that most regulatory interactions are between non-regulatory genes (which for prokaryotes may be a reasonable first-order approximation [18]). These genes form the nodes of the network, while the links between nodes are regulatory interactions. In this case, total gene number $N$ in Eq. 1 is replaced by the number of non-regulatory genes $S$. If we presume that new genes explore outbound regulatory interactions with all existing nodes with equal probability, then this means that the inbound regulatory links to regulated genes are exponentially distributed and described by a random Erdös-Rényi style compact distribution network [22]. Interestingly, it has been shown that the number of transcriptional regulators controlling one gene follows an exponential distribution in *E. coli* [17], as is expected in Erdös-Rényi networks [22]. Existing regulatory networks are of course far from random, but are highly optimized by natural selection. However, if selective forces affect all nodes with equal probability, then the resulting network topology resembles that of a random network.

## Limited complexity of prokaryotes

Regardless of the exact nature of the distributions, the non-linear scaling confirmed in Fig. 1 places prokaryote transcriptional networks firmly in the class of accelerating networks. Regulators are the fastest growing class of proteins [5], and their scaling behaviour has profound implications for the ability of prokaryotes to evolve more complex genetic programs. In particular, the accelerating growth of the regulatory overhead must eventually impose an inherent upper size limit on prokaryote genomes, which we can roughly estimate as being at the point where functional gain is outweighed by regulatory cost, as follows. The total gene number $N$ is composed of both regulatory genes $R$ and non-regulatory genes $S$, and thus $\Delta N = \Delta R + \Delta S$ for any increase in genome size $\Delta N$. In small genomes, growth occurs with the addition of many more non-regulatory than regulatory genes, and $\Delta R << \Delta S$. However, as genomes enlarge, there comes a point where each new non-regulatory gene will be accompanied by the addition of more than one regulatory gene, $\Delta R > \Delta S$. Satisfaction of this constraint roughly indicates when the required regulatory overhead outweighs the gains afforded by additional non-regulatory genes: genome expansion becomes inefficient. From $\Delta R = c\ N\ \Delta N$ (either inferred directly from Fig. 1, or from the derivation of Eq. 1, with $c = 2.40 \times 10^{-5}$ from a fit with slope 2.00), we find that this regime is reached when $c\ N > 1/2$, or $N >$ approximately 20,000 genes. The latter figure is within a factor of two of the observed ceiling of around 10,000 genes in prokaryotes [13]. It may also be noteworthy that the observed limit coincides with the point where the number of operons equals the total number of regulatory interactions affecting them, where the latter is estimated as 5 x the number of regulators [17]. Regardless of the precise limit, the inescapable conclusion stemming from accelerated regulatory network growth is that there is a limit to genome size imposed by regulatory overhead. It may be more than coincidental that the predicted and observed limits are similar.

Our results then suggest that gene numbers and the complexity of prokaryotes may have been constrained by the architecture of their regulatory networks. It is evident that prokaryotes have never reached the size and complexity of multicellular eukaryotes, whose genomes contain 14000-50000 individually regulated protein-coding genes [23-26], and that are subject to alternative splicing to produce many more isoforms, in addition to large numbers of noncoding RNA genes [27]. The low complexity of prokaryotes has previously been attributed to biochemical or environmental factors, e.g., the type of energy metabolism or the absence of particular proteins such as cell-cell signaling proteins or homeobox proteins which are unique to higher eukaryotes. However, such stratagems should have been available to prokaryotes; like eukaryotes, they have had over 4 billion years of evolutionary history in which to explore protein structural and chemical space, aided by lateral gene transfer. It is also often assumed that the increased complexity of eukaryotes is a result of control systems which exploit the increased possibilities afforded by combinatorics of regulatory factors, and of the introduction of new levels of control. However, it would not be difficult to imagine the evolution of larger cis-regulatory regions and new regulatory protein recognition sites in prokaryotic genes. Moreover, the introduction of new levels of control requires the introduction of new regulatory systems and pathways, so the regulatory load problem cannot be avoided in that way.

So how might the developmentally complex eukaryotes have bypassed this constraint? The only general way to relieve the problem is either to reduce the level of connectivity of the regulatory network (which is the opposite of what might be expected in a complex system), or to fundamentally change the nature of $R$, so that the numbers of regulatory factors may be expanded faster than the numbers of regulated proteins. Given that noncoding RNA accounts for about 97% of all transcriptional output in humans [28], and that many complex genetic phenomena in higher organisms are RNA-directed [29], it seems likely that RNA (which is utilized only for a few specific functions in prokaryotes) has been co-opted by the eukaryotes to solve this problem, enabling the programming of large numbers of different cell states and developmental trajectories in complex organisms like humans [30]. The regulatory advantage of RNA is its ability to convey sequence-specific signals (like a zip code or bit string) to receptive targets, while requiring 1.5 orders of magnitude less genomic sequence and correspondingly lower metabolic costs than proteins [29, 31].

# Conclusions

General arguments on the scaling of regulatory networks, as well as two alternative models, predict a faster than linear increase of transcriptional regulatory overhead with gene number. This is confirmed by genomic data. Both models and the empirical data are most consistent with a quadratic growth of transcription regulator number with total gene number. This links transcriptional networks to the emerging field of accelerating networks. The observed non-linear scaling implies a limit on network growth, and therefore on genome size and complexity, within any given regulatory architecture. Our rough prediction of this limit lies within a factor of two of the biologically observed size limit of prokaryotes. This implicates regulatory network structure as a defining feature distinguishing prokaryotes from complex eukaryotes.

# Methods

### Data mining

Profiles in Pfam [16] (http://pfam.wustl.edu) were identified that were DNA or RNA binding and either had known regulatory function or demonstrated sequence specific binding. Pfam was searched by keywords such as "DNA bind", "RNA bind", "regulator" and "transcription factor". In this way almost half of all Pfam profiles were examined (see Supplemental Table 1 for a list of included Pfam profiles). Viral and Eukaryal profiles were included if the profiles fitted the above criteria. TIG immunoglobulin-like domains were excluded as they bind substrates apart from DNA.

Complete, annotated genes were downloaded from NCBI [32]. hmmpfam from Hmmer 2.1.2 was used to identify all proteins that fit any of the selected Pfam profiles. The expectation cutoff for a valid profile match was set at $10^{-4}$. The results were parsed and counted using a Perl script. A list of species, gene numbers, and regulator numbers is provided as Supplemental Table 2. Graphs (not shown) using different definitions of regulatory proteins were also made using COG functional categories, subsets of the Pfam profiles in Supplemental Table 1 and functional classification from genome annotation. All such graphs showed similar behaviours to Figure 1.

## Connectivity

The average connectivity of genes was estimated as follows. If on average each of $R$ regulators connects to five operons, as is the case in *E. coli* [15], then one operon is accessed by $5R/(N_{op}) = 8R/N$ regulators (substituting the asymptotic relationship $N_{op} \approx 0.6 \times N$ [17]. Each of these regulators is also linked to four other operons. Assuming independence of the regulator connections, each gene in the original operon is therefore directly linked to $4 \times 8R/N$ other operons, or $4 \times 8R/(N \times 0.6) = 53R/N$ genes. As a proportion of the total $N$, any one gene is therefore connected to $C = 53R/N^2$ other genes. Substituting $R/N^2$ as estimated from Fig. 1, this gives $C = 0.06\%$ (gene connectivity).

# Acknowledgements

MJL acknowledges financial support by The Wellcome Trust.

**Supplemental Table 1:** Pfam profiles of regulatory proteins showing the name and description of each Pfam profile used to identify transcriptional regulators in each genome

| Name | Description |
| --- | --- |
| ANTAR | ANTAR domain |
| AraC_binding | Arabinose operon regulatory protein |
| Arc | Arc-like DNA binding domain |
| Arg_repressor | Arginine repressor, DNA binding domain |
| ARID | ARID/BRIGHT DNA binding domain |
| ASNC_trans_reg | AsnC family |
| AT_hook | AT hook motif |
| Baculo_IE-1 | Baculovirus immediate-early protein (IE-0) |
| CarD_TRCF | CarD-like/TRCF domain |
| Carla_C4 | Carlavirus putative nucleic acid binding protein |
| CAT_RBD | CAT RNA binding domain |
| crp | Bacterial regulatory proteins, crp family |
| CSD | 'Cold-shock' DNA-binding domain |
| CsrA | Global regulator protein family |
| deoR | Bacterial regulatory proteins, deoR family |
| DM-domain | DM DNA binding domain |
| dsDNA_bind | Double-stranded DNA-binding domain |
| dsrm | Double-stranded RNA binding motif |
| Fe_dep_repress | Iron dependent repressor, N-terminal DNA binding domain |
| filament_head | Intermediate filament head (DNA binding) region |
| FINO | Fertility inhibition protein (FINO) |
| FUR | Ferric uptake regulator family |
| GATA | GATA zinc finger |
| GerE | Bacterial regulatory proteins, luxR family |
| gntR | Bacterial regulatory proteins, gntR family |
| Herpes_ICP4_N | Herpesvirus ICP4-like protein N-terminal region |
| Histone_HNS | H-NS histone family |
| HLH | Helix-loop-helix DNA-binding domain |
| HrcA | HrcA protein C terminal domain |
| HSF_DNA-bind | HSF-type DNA-binding |
| HTH_1 | Bacterial regulatory helix-turn-helix protein, lysR family |
| HTH_10 | HTH DNA binding domain |
| HTH_3 | Helix-turn-helix |
| HTH_4 | Ribbon-helix-helix protein, copG family |
| HTH_5 | Bacterial regulatory protein, arsR family |
| HTH_6 | Helix-turn-helix domain, rpiR family |
| HTH_8 | Bacterial regulatory protein, Fis family |
| HTH_AraC | Bacterial regulatory helix-turn-helix proteins, araC family |
| HTH_psq | helix-turn-helix, Psq domain |
| IclR | Bacterial transcriptional regulator |
| KilA-N | KilA-N domain |
| lacI | Bacterial regulatory proteins, lacI family |
| LexA_DNA_bind | LexA DNA binding domain |
| LytTR | LytTr DNA-binding domain |
| MarR | MarR family |
| merR | MerR family regulatory protein |
| Mga | M protein trans-acting positive regulator (MGA) |
| Mu_DNA_bind | Mu DNA-binding domain |
| myb_DNA-binding | Myb-like DNA-binding domain |
| NusB | NusB family |
| PadR | Transcriptional regulator PadR-like family |
| PAS | PAS domain |
| PC4 | Transcriptional Coactivator p15 (PC4) |
| PRD | PRD domain |
| PurA | PurA ssDNA and RNA-binding protein |
| response_reg | Response regulator receiver domain |
| rrm | RNA recognition motif. (a.k.a. RRM, RBD, or RNP domain) |
| RseC_MucC | Positive regulator of sigma(E), RseC/MucC |
| S1FA | DNA binding protein S1FA |
| SAND | SAND domain |
| SBP | SBP domain |
| SeqA | SeqA protein |
| SfsA | Sugar fermentation stimulation protein |
| Sigma54_activat | Sigma-54 interaction domain |
| sigma54_DBD | Sigma-54, DNA binding domain |
| sigma70_r1_1 | Sigma-70 factor, region 1.1 |
| sigma70_r2 | Sigma-70 region 2 |
| SIS | SIS domain |
| SpoVT_AbrB | SpoVT / AbrB like domain |
| SRF-TF | SRF-type transcription factor (DNA-binding and dimerisation domain) |
| Sua5_yciO_yrdC | yrdC domain |
| Tat | Transactivating regulatory protein (Tat) |
| T-box | T-box |
| TCP | TCP family transcription factor |
| TEA | TEA/ATTS domain family |

| tetR | Bacterial regulatory proteins, tetR family |
|---|---|
| trans_reg_C | Transcriptional regulatory protein, C terminal |
| TrpBP | Tryptophan RNA-binding attenuator protein |
| Vir_DNA_binding | Viral DNA-binding protein, all alpha domain |
| zf-C2H2 | Zinc finger, C2H2 type |
| zf-C2HC | Zinc finger, C2HC type |
| zf-C4 | Zinc finger, C4 type (two domains) |
| zf-Dof | Dof domain, zinc finger |
| zf-NF-X1 | NF-X1 type zinc finger |
| Zfx_Zfy_act | Zfx / Zfy transcription activation region |

**Supplemental Table 2:** Genomic data for bacteria and archaea showing completely sequenced and annotated prokaryotes used for this study, number of protein coding genes and observed regulators. Bacterial species are shown in green, while archaeal species are shown in blue.

| organism | genes | regulatory genes |
|---|---|---|
| *Mycoplasma genitalium* | 480 | 2 |
| *Buchnera aphidicola Sg* | 545 | 7 |
| *Buchnera sp.* | 574 | 9 |
| *Ureaplasma urealyticum* | 611 | 3 |
| *Wigglesworthia brevipalpis* | 611 | 12 |
| *Mycoplasma pneumoniae* | 688 | 2 |
| *Mycoplasma pulmonis* | 782 | 4 |
| *Tropheryma whipplei twist* | 808 | 6 |
| *Rickettsia prowazekii* | 834 | 12 |
| *Chlamydia trachomatis* | 894 | 10 |
| *Chlamydia muridarum* | 916 | 10 |
| *Treponema pallidum* | 1031 | 14 |
| *Mycoplasma penetrans* | 1037 | 16 |
| *Chlamydophila pneumoniae AR39* | 1110 | 10 |
| *Rickettsia conorii* | 1374 | 15 |
| *Thermoplasma acidophilum* | 1478 | 28 |
| *Helicobacter pylori J99* | 1491 | 9 |
| *Thermoplasma volcanium* | 1526 | 26 |
| *Aquifex aeolicus* | 1553 | 39 |
| *Mycobacterium leprae* | 1605 | 43 |
| *Campylobacter jejuni* | 1634 | 27 |
| *Borrelia burgdorferi* | 1637 | 12 |
| *Methanopyrus kandleri* | 1687 | 15 |
| *Streptococcus pyogenes* | 1696 | 78 |
| *Haemophilus influenzae* | 1709 | 63 |
| *Bifidobacterium longum* | 1729 | 78 |
| *Pyrococcus abyssi* | 1765 | 40 |
| *Methanococcus jannaschii* | 1770 | 28 |
| *Thermotoga maritima* | 1846 | 59 |
| *Methanobacterium thermoautotrophicum* | 1869 | 45 |
| *Streptococcus mutans* | 1960 | 111 |
| *Pasteurella multocida* | 2014 | 68 |
| *Neisseria meningitidis MC58* | 2025 | 40 |
| *Streptococcus pneumoniae R6* | 2043 | 80 |
| *Pyrococcus horikoshii* | 2064 | 33 |
| *Pyrococcus furiosus* | 2065 | 39 |
| *Fusobacterium nucleatum* | 2068 | 53 |
| *Streptococcus agalactiae 2603* | 2124 | 102 |
| *Chlorobium tepidum TLS* | 2252 | 34 |
| *Lactococcus lactis* | 2266 | 106 |
| *Clostridium tetani E88* | 2373 | 95 |
| *Archaeoglobus fulgidus* | 2407 | 85 |
| *Staphylococcus epidermidis ATCC 12228* | 2419 | 72 |
| *Thermosynechococcus elongatus* | 2475 | 60 |
| *Thermoanaerobacter tengcongensis* | 2588 | 115 |
| *Halobacterium sp.* | 2605 | 69 |
| *Pyrobaculum aerophilum* | 2605 | 33 |
| *Staphylococcus aureus MW2* | 2632 | 98 |
| *Aeropyrum pernix* | 2694 | 19 |
| *Clostridium perfringens* | 2723 | 121 |
| *Sulfolobus tokodaii* | 2826 | 68 |
| *Xylella fastidiosa* | 2831 | 71 |
| *Listeria monocytogenes* | 2846 | 177 |
| *Corynebacterium efficiens YS-314* | 2950 | 121 |
| *Sulfolobus solfataricus* | 2977 | 67 |
| *Lactobacillus plantarum* | 3009 | 183 |
| *Listeria innocua* | 3043 | 175 |
| *Deinococcus radiodurans* | 3102 | 99 |
| *Synechocystis PCC6803* | 3169 | 111 |
| *Brucella melitensis* | 3198 | 169 |
| *Brucella suis 1330* | 3264 | 164 |
| *Methanosarcina mazei* | 3371 | 76 |
| *Oceanobacillus iheyensis* | 3496 | 183 |
| *Caulobacter crescentus* | 3737 | 237 |
| *Vibrio cholerae* | 3828 | 223 |
| *Clostridium acetobutylicum* | 3848 | 221 |
| *Bacillus halodurans* | 4066 | 252 |
| *Yersinia pestis KIM* | 4090 | 196 |
| *Bacillus subtilis* | 4100 | 247 |
| *Shigella flexneri 2a* | 4180 | 216 |
| *Xanthomonas campestris* | 4181 | 225 |
| *Mycobacterium tuberculosis CDC1551* | 4187 | 160 |
| *Escherichia coli K12* | 4289 | 275 |
| *Xanthomonas citri* | 4312 | 233 |
| *Vibrio vulnificus CMCP6* | 4537 | 308 |
| *Methanosarcina acetivorans* | 4540 | 109 |
| *Salmonella typhimurium LT2* | 4553 | 304 |
| *Leptospira interrogans* | 4727 | 117 |
| *Salmonella typhi* | 4767 | 279 |
| *Shewanella oneidensis* | 4778 | 233 |
| *Ralstonia solanacearum* | 5116 | 345 |
| *Agrobacterium tumefaciens C58* | 5301 | 392 |
| *Pseudomonas putida KT2440* | 5350 | 353 |
| *Pseudomonas aeruginosa PA01* | 5565 | 484 |
| *Nostoc sp.* | 6129 | 267 |
| *Sinorhizobium meliloti* | 6205 | 449 |
| *Streptomyces avermitilis* | 7671 | 617 |
| *Streptomyces coelicolor* | 7897 | 704 |
| *Bradyrhizobium japonicum* | 8317 | 560 |